\documentclass[fleqn,usenatbib]{mnras}

\PassOptionsToPackage{colorlinks=true,linkcolor=blue,citecolor=blue}{hyperref}

\usepackage[compact]{titlesec} 
\titleformat{\section}{\bfseries\uppercase}{\thesection}{1em}{}
\titlespacing*{\section}{0pt}{10pt}{5pt} 

\usepackage{newtxtext,newtxmath}
\usepackage[T1]{fontenc}
\usepackage{enumitem} 

\DeclareRobustCommand{\VAN}[3]{#2}
\let\VANthebibliography\thebibliography
\def\thebibliography{\DeclareRobustCommand{\VAN}[3]{##3}\VANthebibliography}

\usepackage{graphicx}	
\usepackage{amsmath}	
\usepackage{xcolor}	    
\usepackage[abs]{overpic}
\usepackage{booktabs}
\hypersetup{colorlinks=true,linkcolor=blue,citecolor=blue}
\usepackage{orcidlink}

\newcommand{\truth}{\texttt{Foreground-free}}
\newcommand{\avoidance}{\texttt{Avoidance}}
\newcommand{\removal}{\texttt{Removal}}
\newcommand{\TwentyOneEMU}{\texttt{21cmEMU}}
\newcommand{\TwentyOneFAST}{\texttt{21cmFAST}}

\newcommand{\RefereeReply}[1]{\textcolor{black}{#1}}

\newcommand{\addorcid}[1]{\orcidlink{\csname orcidauthor#1\endcsname}}

\newcommand{\figcorner}{Fig.~\ref{fig:corner_all_reds}(a)}
\newcommand{\figwhisker}{Fig.~\ref{fig:whisker}(b)}

\title[Foreground mitigation's impact on constraints]{Impact of 21-cm foreground mitigation strategies on reionization power spectrum constraints}

\author[S. K. Giri \& F. Mertens]{
Sambit K. Giri\addorcid{SG}$^{1}$\thanks{E-mail: sambit.giri@astro.su.se}
and
Florent Mertens\addorcid{FM}$^{2}$\thanks{E-mail: florent.mertens@obspm.fr}
\\
$^{1}$Department of Astronomy and Oskar Klein Centre, AlbaNova, Stockholm University, SE-10691 Stockholm, Sweden\\
$^{2}$LUX, Observatoire de Paris, Universit\'e PSL, Sorbonne Universit\'e, CNRS, 75014 Paris, France
}

\date{Accepted XXX. Received YYY; in original form ZZZ}
\pubyear{\the\year{}}

\begin{document}
\label{firstpage}
\pagerange{\pageref{firstpage}--\pageref{lastpage}}
\maketitle

\begin{abstract}
The 21-cm signal probes the intergalactic medium during the Epoch of Reionization (EoR) but is overwhelmed by astrophysical foregrounds orders of magnitude stronger than the cosmological signal. We evaluate two mitigation strategies: (i) Foreground Avoidance, restricting analysis to the EoR window in Fourier space, and (ii) Foreground Removal via Gaussian Process Regression, which exploits spectral smoothness to statistically separate contaminants and reclaim modes within the contaminated wedge. Both introduce systematic biases of up to $\approx$1$\sigma$ in astrophysical parameters such as the minimum star-forming halo mass and ionising escape fraction, with avoidance posteriors consistently broader than removal owing to its restricted visibility coverage. The global reionization history is recovered within the 95\% credible interval (CI), though the neutral fraction at late reionization epochs shows a persistent bias reflecting the difficulty of its inference from the power spectrum alone. Multi-redshift inference is susceptible to contamination from poorly mitigated bins. Excluding such bins significantly reduces parameter biases, but identifying them requires independent quality metrics. When restricted to identical length scales, both strategies recover similar power spectra, yielding posteriors in similar regions.
\end{abstract}

\begin{keywords}
cosmology: theory -- dark ages, reionization, first stars -- galaxies: high-redshift -- intergalactic medium
\end{keywords}


\section{Introduction}

The redshifted 21-cm hyperfine transition of neutral hydrogen ($\mathrm{H}\text{\small I}$) is a unique probe of the Epoch of Reionization (EoR), encoding both the large-scale structure formation of the high-redshift Universe \citep[e.g.,][]{giri2022imprints,schneider2023cosmological} and the astrophysics of early galaxies \citep[e.g.,][]{hutter2021astraeus,schaeffer2023beorn}. Current radio interferometers such as the Low Frequency Array (LOFAR), the Murchison Widefield Array (MWA), and the Hydrogen Epoch of Reionization Array (HERA) have put upper limits on the 21-cm power spectrum and begun placing constraints on the thermal and ionization state of the intergalactic medium (IGM) \citep[e.g.,][]{mertens2025deeper,nunhokee2025limits,hera2026phaseII}. The Square Kilometre Array Observatory (SKAO) is expected to deliver high signal-to-noise measurements of the 21-cm power spectrum \citep{mellema2013reionization}, making it essential to understand how data-processing choices impact astrophysical inference.

A central challenge is the presence of astrophysical foregrounds that exceed the 21-cm signal by several orders of magnitude. Because these foregrounds are spectrally smooth, their power is largely confined to a wedge-shaped region in the cylindrical $(k_\perp, k_\parallel)$ Fourier plane -- the ``foreground wedge'' -- while the uncontaminated portion is known as the ``EoR window''. Two main mitigation strategies are commonly employed. Foreground \textit{avoidance} strategy restricts the analysis to the EoR window \citep[e.g.,][]{liu2014epoch,pober2016importance}, preserving data fidelity at the cost of losing large-scale modes. In contrast, foreground \textit{removal} strategy models and subtracts foreground emission \citep[e.g.,][]{mertens2018statistical,mertens2024retrieving}, recovering additional modes but dependent on a statistical foreground model. 

Previous 21-cm inference studies have either used idealized foreground-free simulations \citep[e.g.,][]{greig201521cmmc,park2019inferring} or analysed real interferometric data where the astrophysical ground truth is unknown \citep[e.g.,][]{ghara2025constraints}. In both cases, mitigation-induced biases in the recovered posteriors cannot be isolated. A recent SKA Science Data Challenge \citep[SDC3a;][]{bonaldi2025sdc3a} was designed to address this gap. Multiple independent teams applied foreground mitigation pipelines to the same simulated 1000-hour SKA-Low dataset while remaining blind to the ground truth signal during analysis. This blind design eliminated confirmation bias and closely mimicked a real observational campaign, resulting in systematically different recovered power spectra 
whose downstream impact on physical interpretation can be assessed unambiguously.

In this work, we use the SDC3a power spectra to quantify how foreground mitigation strategies, redshift selection, and accessible Fourier modes propagate into reionization parameter constraints. This provides an explicit assessment of the information trade-offs inherent in foreground mitigation during the SKA era. While future SKA data will eventually support more advanced imaging and non-Gaussian statistics \citep[e.g.,][]{giri2018optimal}, the 21-cm power spectrum remains the primary benchmark for the initial EoR science programme. 
This letter is structured as follows: Sec.~\ref{sec:21cm} describes the 21-cm signal and the astrophysical model, Sec.~\ref{sec:fore_mitigation_method} outlines the foreground mitigation strategies, Sec.~\ref{sec:results} presents the parameter inference results, and Sec.~\ref{sec:conclusion} gives our conclusions.

\section{21-cm signal}\label{sec:21cm}

Radio telescopes measure the differential brightness temperature corresponding to the 21-cm signal at position $\pmb{r}$ that is given as,
\begin{eqnarray}
    \delta T_\mathrm{b}(\pmb{r},z) \approx 27 x_\mathrm{HI}(\pmb{r},z)\left[1+\delta(\pmb{r},z)\right] \left(1-\frac{T_\mathrm{CMB}(z)}{T_\mathrm{S}(\pmb{r},z)}\right) \nonumber \\
    \left(\frac{1+z}{10}\frac{0.15}{\Omega_\mathrm{m} h^2}\right)^{1/2}\left(\frac{\Omega_\mathrm{b}h^2}{0.023}\right) \mathrm{mK} \ ,
    \label{eq:dTb}
\end{eqnarray}
where $x_\mathrm{HI}$, $T_\mathrm{S}$, and $\delta$ are the $\mathrm{H}\text{\small I}$ fraction, spin temperature, and gas overdensity. We assume a flat $\Lambda$CDM cosmology with $\{\Omega_\Lambda, \Omega_\mathrm{m}, \Omega_\mathrm{b}, h, \sigma_8, n_\mathrm{s}\} = \{0.69, 0.31, 0.049, 0.68, 0.82, 0.97\}$ \citep{planck2020cosmology}; all lengths are comoving. Next, we describe modelling of astrophysics and signal contaminants. 

\subsection{Astrophysical models}\label{sec:astro_model}

We adopt the parametrization of \citet{park2019inferring}, which was also used to generate the SDC3a ground truth. Star formation efficiency scales as $f_\star(M_h)=f_{\star,10}(M_h/10^{10}M_\odot)^{\alpha_\star}$, the ionising escape fraction as $f_\mathrm{esc}(M_h)=f_\mathrm{esc,10}(M_h/10^{10}M_\odot)^{\alpha_\mathrm{esc}}$, and star formation is exponentially suppressed below a turnover halo mass $M_\mathrm{turn}$ via a duty cycle. We fix $\alpha_\mathrm{esc}=0.043$ as it is degenerate with $\alpha_\star$ for the sources during reionization, leaving four free astrophysical parameters (Table~\ref{tab:prior_ranges}). The star formation timescale $t_\star$ is fixed as it is poorly constrained by the 21-cm power spectrum alone \citep{breitman202421cmemu}, and the X-ray spectral parameters are set to the SDC3a fiducial values---though the total X-ray photon budget remains modulated by $f_{\star,10}$ and $\alpha_\star$. Model predictions are provided by \TwentyOneEMU{} \citep{breitman202421cmemu}, an emulator trained on \TwentyOneFAST{} \citep{mesinger201121cmfast} that delivers 21-cm power spectra and reionization histories across $6\lesssim z \lesssim 21$ and wave-modes $0.05\lesssim k/(h\,\mathrm{Mpc}^{-1})\lesssim 1.63$. \RefereeReply{Since the SDC3a signal was generated with \TwentyOneFAST{}, the mock observations and inference models are consistent.}

\begin{table}
\caption{Prior ranges and SDC3a truth values for the four free astrophysical parameters, and truth values for the neutral fraction ($x_\mathrm{HI}$) at three representative redshifts during reionization (derived quantities; no prior range applies). Priors are flat within the bounds of the \TwentyOneEMU{} training space.}
\label{tab:prior_ranges}
\centering
\begin{tabular}{c c c}
\hline\hline
Parameter & Prior range & Ground truth \\
\hline
    $\log_{10}(M_\mathrm{turn}/M_\odot)$ & $[7.0, 10.0]$ & 8.277 \\
    $\log_{10}(f_\mathrm{\star,10})$ & $[-3.0, -1.0]$ & $-1.509$ \\
    $\alpha_\mathrm{\star}$ & $[0.0, 1.0]$ & 0.496 \\
    $\log_{10}(f_\mathrm{esc,10})$ & $[-2.0, 0.0]$ & $-1.046$ \\
\hline
    $x_\mathrm{HI}(z=6.54)$ & -- & 0.33 \\
    $x_\mathrm{HI}(z=7.19)$ & -- & 0.58 \\
    $x_\mathrm{HI}(z=7.96)$ & -- & 0.72 \\
\hline
\end{tabular}
\end{table}

\subsection{Signal contamination}\label{sec:signal_contamination}

The 21-cm signal is buried beneath astrophysical foregrounds that exceed it by 3–5 orders of magnitude. These foregrounds are spectrally smooth in principle, but instrumental effects such as antenna beam chromaticity, calibration residuals, and ionospheric phase errors corrupt their spectral structure, complicating the signal separation procedure. Critically, baseline chromaticity couples foreground power into the line-of-sight Fourier modes that constitute the EoR window, contaminating the very scales essential for 21-cm detection. The SDC3a simulations \citep{bonaldi2025sdc3a,bonaldi2026ska} were generated using the \texttt{OSKAR} software package and incorporate these effects---including imperfect de-mixing of bright out-of-field sources, direction-independent gain calibration errors, and thermal noise rescaled to a 1000 h SKA-Low integration---providing a realistic test bed for foreground mitigation strategies.

\section{Foreground mitigation and data recovery}\label{sec:fore_mitigation_method}

In this study, we evaluate two distinct mitigation philosophies---Fourier-space \textit{avoidance} and statistical \textit{removal}---using pipelines submitted by the DOTSS team to the SDC3a \citep{bonaldi2025sdc3a}. 
Below, we briefly describe these two mitigation methods and present the reconstructed 21-cm power spectra from these pipelines.

\subsection{Foreground avoidance}
Foreground avoidance exploits the fact that the smooth spectral nature of astrophysical foregrounds causes their power to be confined to a geometric region of the cylindrical $(k_\perp, k_\parallel)$ Fourier plane known as the ``foreground wedge'' \citep{liu2014epoch,pober2016importance}. Instrumental chromaticity and the telescope's field of view define the extent of this contaminated region. By restricting analysis to the comparatively clean ``EoR window'' using the \textit{DOTSS-21cm Avoidance} pipeline (\avoidance{}), this strategy provides a robust estimate that bypasses the need for statistical foreground--signal separation, thereby removing the risk of model misspecification. However, this data purity comes at the cost of losing information on the largest spatial scales ($k \lesssim 0.2\,h\,\mathrm{Mpc}^{-1}$). 

\subsection{Foreground removal}
Foreground removal attempts to reclaim high-sensitivity modes within the wedge by modelling and subtracting the foreground emission. The \textit{DOTSS-21cm Advanced ML-GPR} pipeline (\removal{}) 
utilizes Gaussian Process Regression (GPR), which exploits the distinct spectral coherence scales of the astrophysical foregrounds relative to the 21-cm signal. By employing a machine-learning-trained kernel to model the frequency-dependent foreground covariance along each line of sight, the pipeline predicts and subtracts the smooth components \citep{mertens2024retrieving,acharya202421}. While this approach in principle recovers the full set of Fourier modes,
it introduces the risk of systematic biases if the statistical model is misspecified. Such biases are particularly evident at redshifts where the 21-cm signal exhibits complex spectral structure that the model may mistake for foreground components.

\subsection{Recovered 21-cm power spectra}\label{sec:recovered_21cm}

\begin{figure*}
\centering
\includegraphics[width=0.85\textwidth,trim={0.25 0.25cm 0.20cm 0.1},clip]{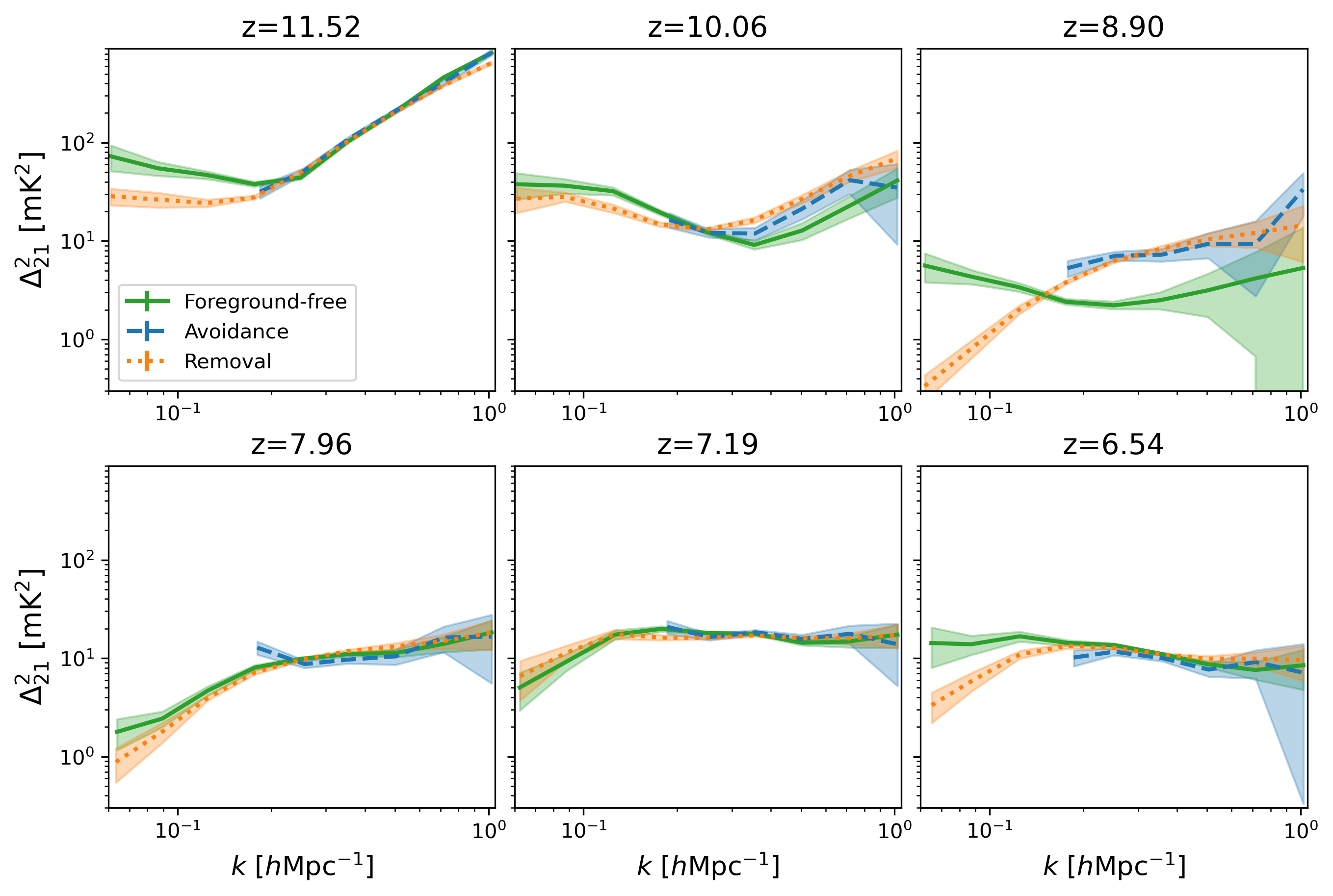}
\vspace{-0.8em}
\caption{
Power spectra ($\Delta^2_\mathrm{21}$) at different redshifts ($z$) showing the \RefereeReply{foreground-free} signal (green solid curves), the reconstruction after foreground avoidance (blue dashed curves), and the reconstruction after foreground removal (orange dotted curves). Shaded regions represent the combined uncertainty from sample variance and instrumental noise. Foreground avoidance does not recover the largest spatial scales ($k \lesssim 0.2\,h\,\mathrm{Mpc}^{-1}$). Both mitigation methods struggle to accurately retrieve the signal at $z \approx 8.90$. 
}
\label{fig:ps_mitigated}
\end{figure*}

Fig.~\ref{fig:ps_mitigated} shows the dimensionless power spectra $\Delta^2_{21}(k)$ at the six redshift bins. \RefereeReply{The shaded regions show the combined uncertainty from sample variance and instrumental noise. Comparing to the \truth{} case,} \avoidance{} consistently loses sensitivity at the largest scales ($k \lesssim 0.2\,h\,\mathrm{Mpc}^{-1}$). Both mitigation methods fail most severely at $z \approx 8.90$, where the reconstructions converge with each other but depart significantly from the \truth{} \RefereeReply{case}. At this epoch, the simultaneous onset of X-ray heating and early reionization produces complex signal power that is difficult to distinguish from foreground residuals, leading to excess power in the reconstruction at $k\gtrsim 0.2\,h\,\mathrm{Mpc}^{-1}$ with a drop at the largest scales accessed by \removal{}.
At $z=6.54$, both methods exhibit a systematic large-scale ($k \lesssim 0.3\,h\,\mathrm{Mpc}^{-1}$) bias, most pronounced in the \removal{} case due to the reclaimed largest scales. In contrast, at the highest redshift bin ($z\approx 11.52$), both methods reconstruct the signal remarkably well. While \removal{} shows a bias in the largest reclaimed scale, it correctly recovers the shape. At $z=7.19$ and $7.96$, both strategies track the truth within $1$--$2\sigma$, with \removal{} successfully reclaiming the large-scale modes. At $z=10.06$, both methods show moderate biases driven by heating-onset spectral structure, but the difference between the reconstruction and the truth is substantially less severe than at $z=8.90$, allowing the bin to retain useful constraining power. Thus, these four better-recovered bins ($z\approx 7.19$, $7.96$, $10.06$, and $11.52$) define our \textit{Best redshifts} scenario, which is explored later.

\begin{figure*}
\makebox[\textwidth][l]{{\large\textbf{(a)} Full posterior distributions --- \textit{All redshifts} scenario}}
\begin{flushleft}
\begin{overpic}[width=0.6\textwidth,trim={0 0.25cm 0 0.1},clip]{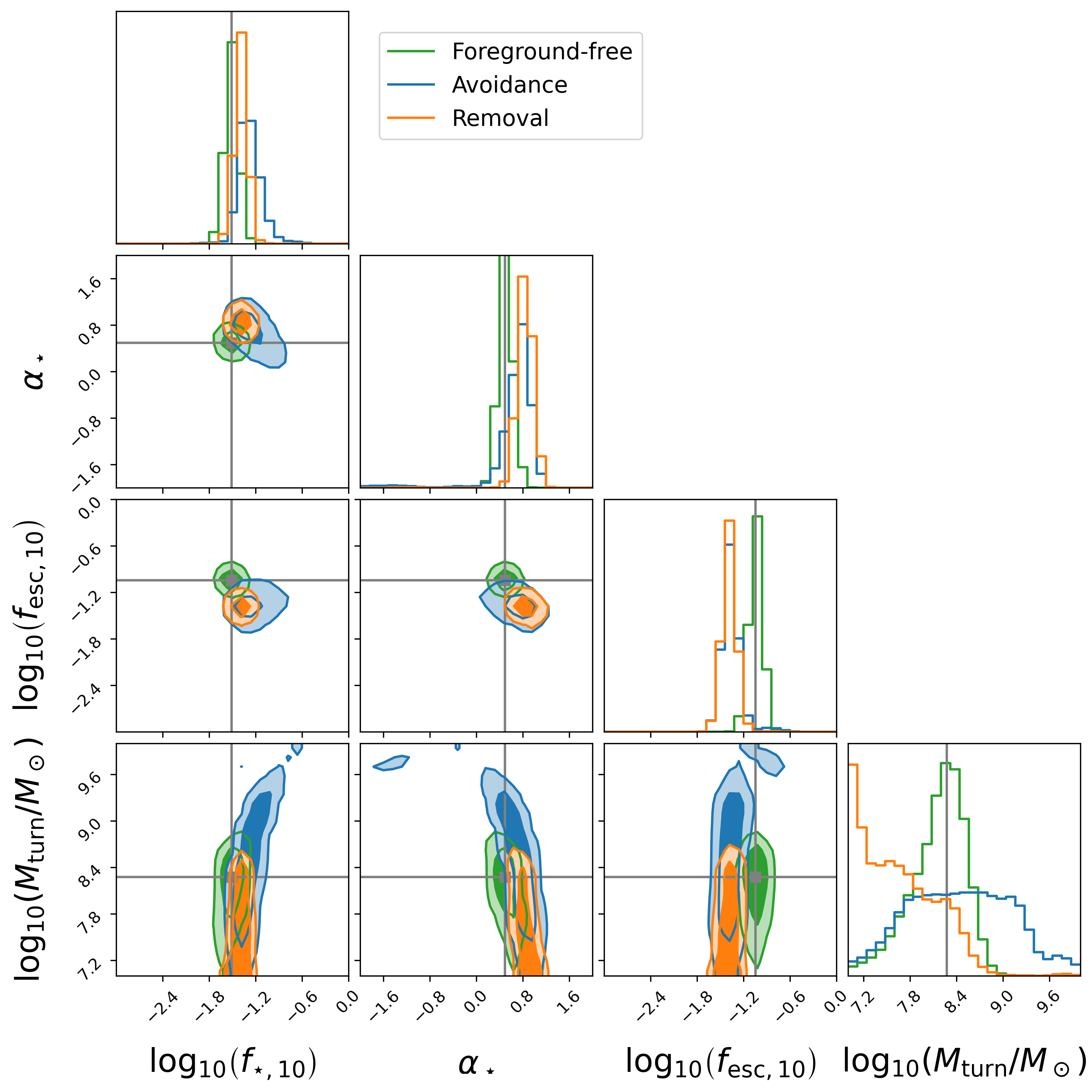}
     \put(283,95){\includegraphics[width=0.41\textwidth]{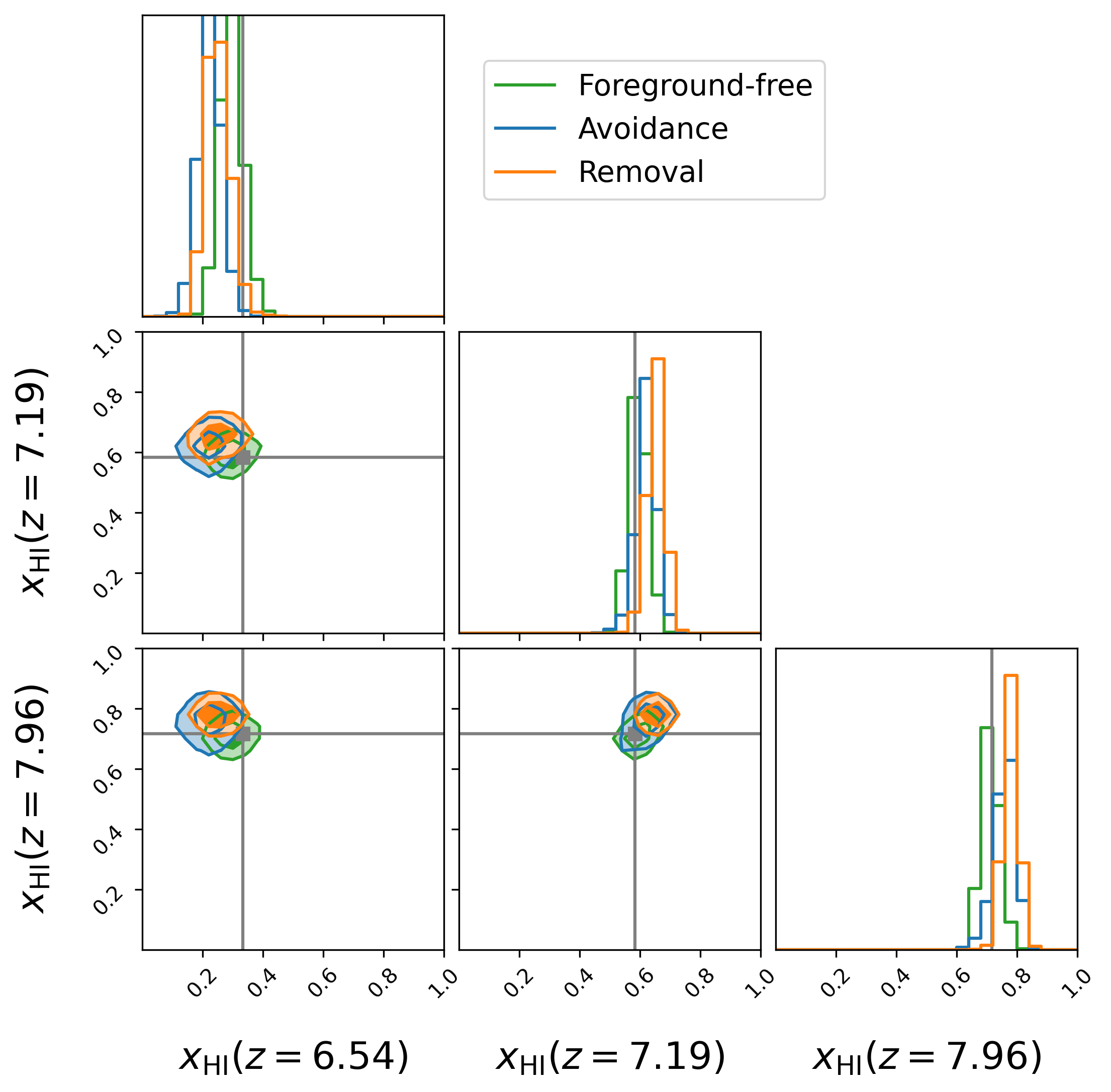}}
\end{overpic}
\end{flushleft}
\vspace{1mm}
\makebox[\textwidth][l]{{\large\textbf{(b)} Marginalized constraints for all scenarios}}
\centering
\includegraphics[width=0.90\textwidth,trim={0 0.25cm 2.5cm 0},clip]{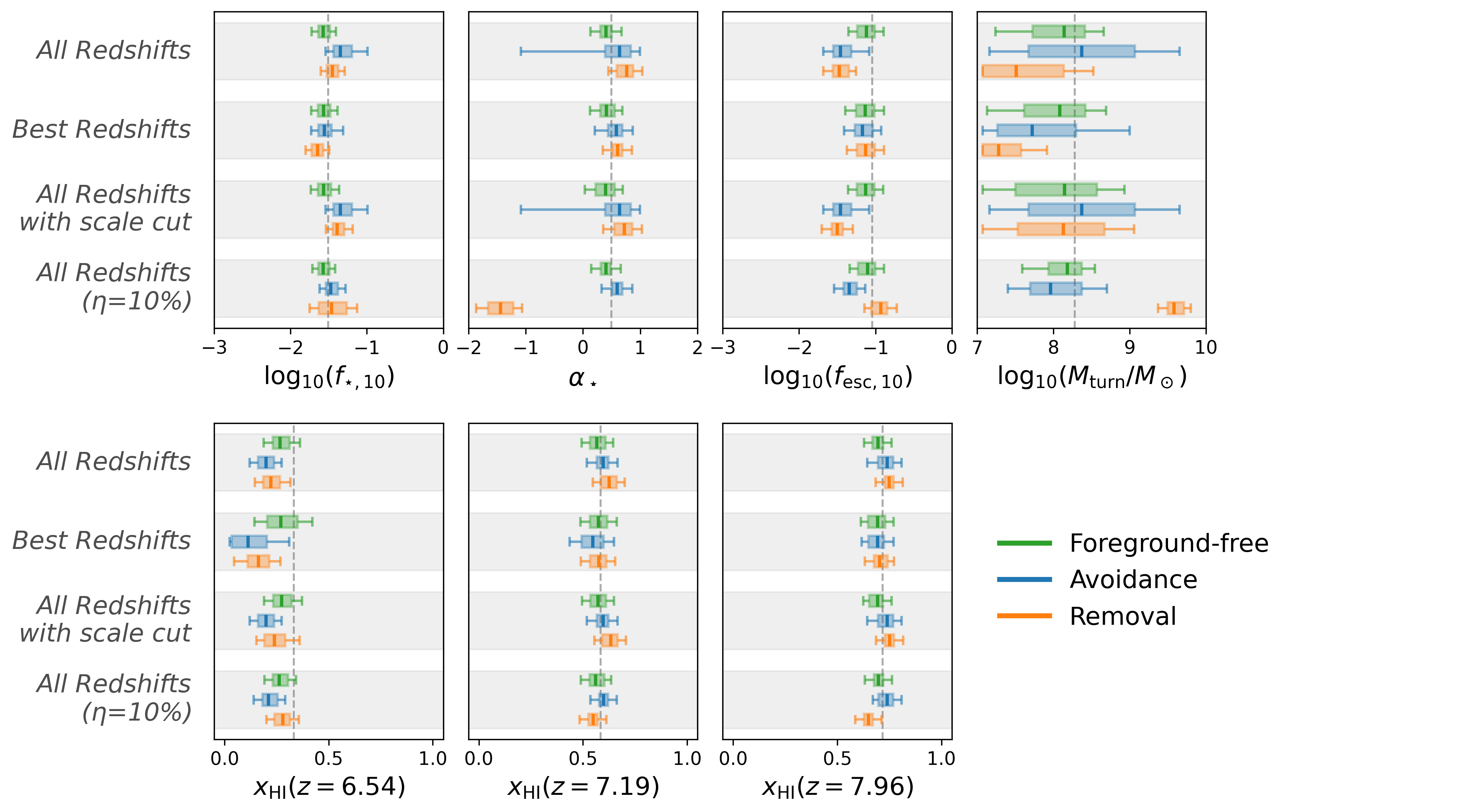}
\vspace{-0.6em}
\caption{
\textbf{(a)} Posterior distributions for the astrophysical and IGM parameters using all six redshift bins (\textit{All redshifts} scenario). The main plot shows constraints on the astrophysical model parameters, where $\log_{10}(M_\mathrm{turn}/M_\odot)$ exhibits the largest bias when using foreground-mitigated signals. Dark and light shaded regions represent the $1\sigma$ and $2\sigma$ credible contours for the \truth{} (green), \avoidance{} (blue), and \removal{} (orange) cases. The inset (top-right) displays constraints on the neutral fraction $x_\mathrm{HI}$ at three representative epochs.
\textbf{(b)} 1D marginalized constraints (68\% and 95\% credible interval, CI) on the astrophysical parameters and $x_\mathrm{HI}$ for four inference scenarios: \textit{All redshifts} (all six bins, $z = 6.54$--$11.52$), \textit{Best redshifts} (four best-quality bins, excluding $z = 8.90$ and $6.54$), \textit{All redshifts with scale cut} (all six bins restricted to $k$-modes accessible to \avoidance{}), and \textit{All redshifts} ($\eta=10\%$) (all six bins with modelling uncertainty reduced to $\eta=10\%$). Vertical dashed lines mark the ground truth values.
}
\label{fig:corner_all_reds}
\label{fig:whisker}
\end{figure*}

\section{Results}\label{sec:results}
We perform Bayesian inference using a Gaussian likelihood with diagonal covariance. The SDC3a power spectra are provided with $\Delta\ln k \approx 0.35$, and we assume no correlations between bins. The variance $\sigma^2_{21}(k,z)$ includes instrumental noise, sample variance, and a modelling uncertainty $\eta\Delta^2_\mathrm{model}$, 
which are added in quadrature. \RefereeReply{Unless specified otherwise, we assume $\eta=20\%$, which is motivated by the differences seen in code comparison studies \citep[e.g.,][]{hutter2018accuracy}}. Model predictions are provided by \TwentyOneEMU{}, with flat priors over the ranges in Table~\ref{tab:prior_ranges}. We sample the posterior using \texttt{nautilus} \citep{lange2023nautilus}, a dynamic nested sampler that efficiently explores highly degenerate parameter spaces. Derived IGM quantities, namely the neutral fraction $x_\mathrm{HI}(z)$, are obtained by propagating posterior samples through \TwentyOneEMU{}.
In the following subsections, we evaluate the impact of foreground mitigation on parameter inference.

\subsection{Results from full redshift range}\label{sec:inference_results}

\figcorner{} shows the full posterior distributions for the \textit{All redshifts} scenario. \RefereeReply{The \truth{} case corresponds to inference on the foreground-free power spectra and serves as the ideal reference.} Both mitigation strategies show qualitatively similar bias patterns. Owing to information loss from the excluded wedge modes, the \avoidance{} posteriors are consistently broader than those of \removal{} across all parameters. The most significant bias is in $\log_{10}(M_\mathrm{turn}/M_\odot)$: the \avoidance{} posterior broadens toward high values while \removal{} is pushed to the lower prior boundary at $7.0$---below which star formation is physically implausible 
\citep{nebrin2023starbursts}. 
The stronger biases in \removal{} are caused by the suppressed large-scale power in the recovered spectra (Fig.~\ref{fig:ps_mitigated}), as discussed further in Sec.~\ref{sec:diagnostic}. 
For both strategies, $\alpha_\star$ is biased slightly high (within the 68\% CI) and $\log_{10}(f_\mathrm{esc,10})$ is biased toward low values. 

The $x_\mathrm{HI}$ at $z=7.96$ and $z=7.19$ (early and middle stages of reionization) are recovered within the 95\% CI for both strategies. $x_\mathrm{HI}$ is less sensitive to mitigation-induced spectral distortions than the source parameters, as $x_\mathrm{HI}$ is primarily encoded in the overall power spectrum amplitude rather than its detailed shape. At $z=6.54$, however, $x_\mathrm{HI}$ shows a persistent $1$--$2\sigma$ bias toward lower values across all scenarios, likely because ionised bubble overlaps at late stages of reionization introduce non-linearities that decouple the power spectrum amplitude from $x_\mathrm{HI}$ \citep[e.g.,][]{georgiev2022large}. 

\subsection{Diagnostic scenarios: Redshift selection, scale sensitivity, and modelling uncertainty}\label{sec:diagnostic}

\figwhisker{} investigates the sources of bias in the \textit{All redshifts} results by comparing the 1D posteriors from three diagnostic scenarios. First, we consider the \textit{Best redshifts} case, where the two most contaminated bins ($z=8.90$ and $6.54$, identified in Sec.~\ref{sec:recovered_21cm}) are excluded, retaining only the four better-recovered bins. This yields a significant improvement in the \removal{} posteriors, with biases in most parameters reduced to approximately $0.5\sigma$. The $\log_{10}(M_\mathrm{turn}/M_\odot)$ remains notably biased even in this best-case scenario, suggesting it is the parameter most sensitive to any residual differences in the mitigated power spectra. The $x_\mathrm{HI}$ posteriors at $z=7.19$ and $7.96$ now peak at the truth values for both strategies. At $z=6.54$, excluded from the likelihood, the $x_\mathrm{HI}$ posterior is instead propagated through the source model from parameters constrained at higher redshifts, and again peaks below the truth.

In the second scenario (\textit{All redshifts with scale cut}), we restrict \removal{} to the same $k$-modes as \avoidance{}, bringing the astrophysical parameter posteriors to similar regions as \avoidance{}. The \avoidance{} posteriors remain somewhat broader even at identical mode coverage, owing to the larger power spectrum variance from the reduced number of visibilities used in the EoR window. This near-convergence confirms that it is the additional $k$-modes recovered by \removal{} that primarily drive the divergence between strategies. The $x_\mathrm{HI}$ constraints at $z=7.19$ and $7.96$ are similar for both strategies. At $z=6.54$, $x_\mathrm{HI}$ remains biased toward lower values for both strategies, though scale-cut \removal{} recovers the truth within the 95\% CI while \avoidance{} does not.

The third diagnostic considers a lower modelling uncertainty ($\eta=10\%$) applied to all inference cases. This sharpens the likelihood, making the inference more sensitive to the inaccurate large-scale power spectra recovered at $z=6.54$ and $z=8.90$. The \avoidance{} posteriors remain centred at similar positions but with narrower credible intervals. In contrast, \removal{} shows markedly larger biases: $\alpha_\star$ is driven $>3\sigma$ toward low values and $\log_{10}(M_\mathrm{turn}/M_\odot)$ is pushed $>3\sigma$ toward high values. Despite these severe source parameter offsets, the $x_\mathrm{HI}$ constraints remain within the 95\% CI, demonstrating the relative resilience of the reionization history. This is consistent with the general resilience of $x_\mathrm{HI}$ to mitigation-induced spectral distortions discussed in Sec.~\ref{sec:inference_results}.

\section{Conclusions}\label{sec:conclusion}

In this work, we assessed how foreground mitigation strategies impact astrophysical constraints derived from 21-cm power spectrum measurements. Our findings are summarized as follows:
\begin{itemize}[noitemsep, topsep=0pt, leftmargin=*]
    \item \textbf{Mitigation-induced astrophysical biases}: Both strategies introduce systematic biases in astrophysical parameters. The most significant impact is on $\log_{10}(M_\mathrm{turn}/M_\odot)$: \avoidance{} posteriors broaden toward high values while \removal{} is pushed to low values. Both strategies also show mild biases in $\alpha_\star$ and $\log_{10}(f_\mathrm{esc,10})$, driven by poorly recovered low-$k$ modes.
    \item \textbf{Resilience of reionization history constraints}: $x_\mathrm{HI}$ shows smaller biases than the source parameters and is recovered within the 95\% CI across all scenarios, because it is primarily encoded in the overall power spectrum amplitude (reasonably well recovered by both strategies) and because degenerate source configurations can yield qualitatively similar ionisation morphologies. The exception is the late reionization epoch ($z=6.54$), where $x_\mathrm{HI}$ shows a persistent bias, likely reflecting non-linearities from ionised bubble overlaps that decouple the power spectrum amplitude from $x_\mathrm{HI}$.
    \item \textbf{Multi-redshift inference}: Multi-redshift inference is susceptible to contamination from poorly recovered bins, which can degrade the joint posterior. In real observations without a known ground truth, identifying such bins requires independent quality metrics. For example, cross-correlations with complementary datasets, such as Lyman-$\alpha$ forest, CMB lensing and galaxy surveys, offer one route \citep[][]{giri2025mapping, hutter202621cm}.
    \item \textbf{Strategy convergence on the power spectrum}: When restricted to identical length scales, both strategies yield posteriors in similar regions, though avoidance retains broader constraints due to the larger power spectrum variance from its restricted visibility coverage. The relative merits of the two strategies are, however, observable-dependent. For example, image-based statistics require sky-plane access that only foreground removal provides.
\end{itemize}
Our analysis is performed within a particular astrophysical modelling framework, and extensions to a broader parameter space or alternative reionization models are important directions for future work. Foreground mitigation remains a primary challenge for 21-cm cosmology, and strategies optimal for the power spectrum need not be optimal for other observables. Modern machine-learning methods can recover contaminated modes beyond the reach of standard mitigation strategies \citep[e.g.,][]{gagnon2021recovering,bianco2025deep}, and higher-order summary statistics \citep[e.g.,][]{cerardi2025implicit,sun2025lfi21cmforest} can complement power-spectrum constraints by recovering information on scales excluded by avoidance. A promising route to reduce these biases is to incorporate foreground models directly into the inference forward model, enabling joint astrophysical and foreground parameter estimation \citep[e.g.,][]{kern2025differentiable}.

\section*{Acknowledgements}
We thank the Square Kilometre Array Observatory (SKAO) for organising the Science Data Challenge (SDC), which has significantly contributed to equipping research teams worldwide for the analysis of upcoming SKA data. We also thank Daniel P. Meerburg, Garrelt Mellema and Benoit Semelin for useful discussion, and the anonymous reviewer for their constructive comments.
SKG acknowledges support from Olle Engkvist Stiftelse (grant no. 232-0238).

\section*{Data Availability}
The data underlying this article will be shared on reasonable request to the corresponding author.


\bibliographystyle{mnras}
\bibliography{references}


\bsp
\label{lastpage}
\end{document}